\documentstyle[11pt,IAU207_pasp,twoside,psfig]{article}
\def\edcomment#1{\iffalse\marginpar{\raggedright\sl#1\/}\else\relax\fi}
\reversemarginpar 

\begin{document} \title{Triggering the Formation of Young Clusters}
\author{Bruce G. Elmegreen} \affil{IBM Research Division, T.J. Watson
Research Center, PO Box 218, Yorktown Hts., NY 10598 USA} 

\begin{abstract} Star formation is triggered in essentially three ways:
(1) the pressures from existing stars collect and squeeze nearby dense
gas into gravitationally unstable configurations, (2) random compression
from supersonic turbulence makes new clouds and clumps, some of which
are gravitationally unstable, and (3) gravitational instabilities in
large parts of a galaxy disk make giant new clouds and spiral arms that
fragment by the other two processes into a hierarchy of smaller
star-forming pieces. Examples of each process are given. Most dense
clusters in the solar neighborhood were triggered by external stellar
pressures. Most clusters and young stars on larger scales are organized
into hierarchical patterns with an age-size correlation, suggestive of
turbulence. Beads-on-a-string of star formation in spiral arms and
resonance rings indicate gravitational instabilities. The turbulence
model explains the mass spectrum of clusters, the correlation between
the fraction of star formation in the form of clusters and the star
formation rate, found by Larsen \& Richtler, and the correlation between
the size of the largest cluster and the number of clusters in a galaxy.
\end{abstract}

to appear in "IAU Symp. 207 on Extragalactic Star Clusters,"
eds. E. Grebel, D. Geisler, \& D. Minniti, ASP Conference series, 
2001, in press (held in Pucon Chile, March 12-16, 2001).

\section{Introduction} 

Star formation may be triggered in a variety of ways. 
This review concentrates on three mechanisms:
(1) sequential, in which pressures from one generation of stars move and
compress the surrounding gas, causing another generation of stars to
form, (2) turbulent, in which random and chaotic supersonic flows
converge and diverge, bringing the gas into dense regions that last for
about a crossing time, and (3) self-gravitational, in which density
perturbations grow as a result of gravitational forces. Other
compressive instabilities may contribute to the gravitational
instability to drive star formation, such as the thermal or Parker
instabilities. Cloud collisions are included in (2) and (3), considering
that most clouds are part of a pervasive fractal network that is
probably generated by turbulence and self-gravity. Density wave
triggering is included in (3) if the waves induce local gravitational
instabilities, forming the characteristic beads-on-a-string pattern of
giant cloud complexes in spiral arms. 

These processes trigger both individual stars and whole star clusters,
the difference being a matter of scale for the self-gravitating cloud
that forms. If a cloud is big enough, then smaller versions of the same
three processes can trigger smaller clouds inside of it, eventually
getting down to the scale of individual stars. For example, the dust
lane in a spiral arm may be gravitationally unstable to make $10^7$
M$_\odot$ cloud complexes. These complexes will be born with a lot of
turbulent energy, from both the pre-cloud gas and the binding energy, and
the turbulence will randomly compress the gas inside, making smaller
clouds. If some of these clouds form stars, then the pressures caused by
these stars can trigger more star formation inside other
turbulence-compressed clouds. Thus all three processes can happen
simultaneously in the same region, sometimes in a nested fashion and
sometimes in juxtaposition. The point of this classification of
processes is not to suggest that star formation follows only one of
three possible paths, but to distinguish between the various
morphologies of clustering that we see in young stellar regions. 

Many young clusters are embedded in dense gas clumps that are at the
interface between a molecular cloud and a high-pressure source, such as
an HII region. These clusters were probably triggered by the HII region
as it expanded into the cloud and are examples of the first process.
When the same clusters are viewed from a greater distance, they are
often found to be part of a fractal hierarchy of many other clusters,
including slightly older OB associations and even older star complexes.
This hierarchical pattern and a corresponding correlation between age
and size has the same form as in a compressibly turbulent fluid, in
which case the turbulent nature of cluster triggering becomes evident on
the larger scale. It might also be true that all of this region is one
of the beads in a galactic spiral arm, or perhaps it is a flocculent
spiral arm by itself. We know that gaseous self-gravity is important on
these large scales, so the whole process will appear to have begun with
an instability that led to a cascade of interconnected events. What is
the actual process of cluster formation in a case like this (which
resembles Gould's Belt)? Is it (1), (2), or (3) in the list above? 
The individual clusters
resulted from a combination of processes. Yet the distinct morphologies
of where and when they formed help us to understand which process
dominates on which scale. 

In the following sections, observations of these three processes are
reviewed. A more comprehensive review of the sequential process is in
Elmegreen (1998). Other reviews of star formation are in Evans (1999)
and in the conference proceedings for Protostars and Planets IV (Univ.
Arizona Press, 2000). 

\section{Sequentially Triggered Star Formation} 

There are many examples in the recent literature where clusters appear
to have been triggered by nearby high-pressure events. Lefloch \&
Chernicharo (2000) found 12 $\mu$ cores that are probably protostars
inside dense clumps that are seen in 1.3 mm at the edge of the Trifid
Nebula. They derived an expansion age of the HII region to be 0.35 My,
and a preshock density $2\times10^3$ cm$^{-3}$. The instability time,
$0.25\left(G\rho\right)^{-1/2}\approx0.35$ My (Elmegreen 1989a), agrees
with the expansion age, suggesting triggering. A similar
configuration with an embedded cluster next to the Rosette nebula was
discussed by Phelps \& Lada (1997). 

Lada et al. (1991) found two dense embedded clusters in the Orion
molecular cloud at the locations of the densest cloud cores. These cores
are the heads of giant cometary clouds, pointing toward and probably
formed by the expansion of the Orion nebula (Bally et al. 1987). The
head of the southern core is a thin cloud running parallel to the
ionization front (Dutrey, et al. 1991) that is probably compressed gas.
Many young stars and protostars are lined up along this strip (Reipurth,
Rodriguez \& Chini 1999). Images from 2MASS of the whole region are in
Carpenter (2000). 

An IRAS image of the embedded cluster near rho Ophiuchus makes it look
like a comet head too (IPAC image from http://www.ipac.caltech.edu). The
source of pressure is in the northwest. This is the triggering
configuration proposed by de Geus (1992) who attributed the compression
to shocks from the Upper Sco OB association. Shocks from the Cen-Lupus
association probably triggered Upper Sco in a previous step (Preibisch
\& Zinnecker 1999). Another giant shell containing several young
clusters surrounds the HII region and bubble source W5, as seen in IRAS
15 $\mu$ and 25 $\mu$ maps (Kerton \& Martin 2000). 

One of the first examples of sequential star formation, found by Sancisi
et al. (1974), includes the Per OB2 association and its two clusters, IC
348 and NGC 1333. Sancisi et al. pointed out that this association lies
in an OH+HI shell with a peculiar velocity, and they suggested that the
clusters were triggered by the shell's expansion. CO maps of the Perseus
region by Sargent (1979) found other age sequences inside the
association. The embedded clusters are not very dense (Lada \& Lada
1995; Lada, Alves, \& Lada 1996), but the region is older than Orion
(Palla \& Stahler 1999, 2000) and the pressures are not as large. Now
the whole region can be seen with 2MASS (Carpenter 2000). 

There are many other embedded clusters in the solar neighborhood. Most
of them look triggered because of their proximity to high pressure
sources (see the table of sources and discussion in Elmegreen et al.
2000). Clusters whose formation is not so clear tend to be older, so
their pressure sources could have decayed. The partially embedded
cluster IC 5146 is an example (Lada, Alves, \& Lada 1999). It has no
obvious source of high pressure nearby, but it is at the tip of an
elongated cloud that makes it look triggered anyway. 

Supernovae can trigger star formation, but most catalogued supernova
remnants are too young to have started this process yet. A recent
example of supernova triggering seems to be G349.7+0.2, which has 3 IRAS
sources along the perimeter of a supernova shell (Reynoso \& Magnum
2001). Other shells have triggered star formation on their periphery
(Xie \& Goldsmith 1994; Yamaguchi et al. 2001), but these shells are
probably older than single supernova remnants and result from a
combination of supernova and stellar wind pressures in an aging OB
association (McCray \& Kafatos 1987). 

Giant shells in other galaxies are often seen with triggered star
formation along their peripheries (Brinks \& Bajaja 1986; Puche et al.
1992; Wilcots \& Miller 1998; Steward et al. 2000; Stewart \& Walter
2000). IC 2574 (Walter \& Brinks 1999; Steward \& Walter 2000) has an
example where HII regions on a shell contain clusters $\sim3$ My old,
and another cluster 11 My old is in the center. 

What are the characteristics of sequentially triggered star formation?
To be reasonably sure that such triggering happened, we need two
distinct regions of star formation observable as either stars or IR
clusters with a separation and age difference that has a ratio equal to
a reasonable propagation speed. The age difference should also be
several tenths of the ambient gas dynamical time,
$\left(G\rho_0\right)^{-1/2}$, for density $\rho_0$ near the younger
cluster. Between the two clusters, there should be a region with a low
density where the gas was cleared away by the pressure disturbance from
the older cluster. The clusters should also be young enough that neither
has had time to move very far from its point of origin. 

Observations of star formation within the nearest several kiloparsecs
suggest that most of it occurred in dense clusters, and that most of
these clusters were triggered by previous generations of stars. Further
from the Sun, such triggering will be harder to see unless there is an
obvious shell. This is because most triggering inside OB associations has
a short length scale, perhaps 1 to 10 pc, and external galaxies are too
far away to resolve this scale. Also, shells tend to shear away quickly
and high pressures last only a relatively short time, so it will be
unusual to catch a triggering event in the act. Most clusters also
disperse rather quickly. Only 10\% of young local stars are in clusters, even
though most stars probably formed this way, and this suggests that most newborn
clusters lose their stars quickly. A triggering act has to be confirmed
while the stars are still in their clusters so we can be sure where they
formed. 

\section{Turbulence Triggering} 

Supersonic turbulence compresses gas in random places, making transient
clouds that last for about an internal crossing time once they form --
longer if we consider also their formation time, which is the flow time in
the external medium (Elmegreen 1993). Numerical simulations show this
process well (V\'azquez-Semadeni, Passot, \& Pouquet 1996; 
Ballesteros-Paredes, V\'azquez-Semadeni, \& Scalo 1999; 
Klessen, Heitsch, \& Mac Low 2000; Padoan et al. 2001). 

A characteristic of turbulence triggering is that clusters are born in
hierarchical and fractal patterns on both galactic scales (Elmegreen \&
Elmegreen 2001) and local scales.  T Tauri stars in the local 
field are hierarchically distributed 
(Gomez, et al. 1993), and the embedded stars in clusters can be
hierarchical too (Elmegreen
2000; Testi et al. 2000). Hierarchical structure is probably the
result of the scaling between velocity and separation for a turbulent
fluid: large scales have large relative velocities, making large
compressed sub-regions, and small scales have small relative velocities,
making small compressed sub-regions. 

Interstellar turbulence is pervasive, although not all regions are
turbulent and not all structures are fractal. Non-fractal structures
include expanding shells, two-arm density wave modes, relaxed clusters,
cometary clouds and Bok globules. Fractal structures include diffuse
clouds, flocculent spiral arms, young clusters where the stars have not
moved much from their birth, and the interiors of many weakly
self-gravitating clouds. When directed pressures shape a cloud, shell
or spiral arm, the overall morphology is determined by those pressures,
but when turbulence is free to define a structure independent of rigid
boundaries, fractal patterns appear.

Turbulence can be generated by expansion or other directed pressures as
the moving gas mixes with the environment and undergoes Rayleigh-Taylor
and Kelvin-Helmholtz instabilities. Self-gravity can make turbulence
as gas collapses locally and mixes. It can also be generated by Parker
(1965) instabilities (Asseo et al. 1978) and Balbus-Hawley (1991)
instabilities (Sellwood \& Balbus 1999) which involve magnetic mixing.
Turbulence decays rapidly (Stone, Ostriker, \& Gammie 1998; MacLow, et
al. 1998), but in a self-gravitating medium, there is always a source
for more turbulence as the region contracts.

Star formation in a turbulent medium should have several special
characteristics. First, the star-forming clouds should be randomly placed
in part of an overall fractal gas distribution. This fractal gas will
have a scale-free power spectrum (Crovisier \& Dickey 1983; Green 1993;
Lazarian \& Pogosyan 2000; St\"utzki et al. 1998; Stanimirovic et al.
1999; Westpfahl et al. 1999; Elmegreen, Kim, \& Staveley-Smith 2001).
The young stars that form in these clouds should have the same fractal
pattern as the gas at the time of their birth.  Their birth times could
be random, although there could be some fractal substructure in their
birth times too. The clouds and clumps in a fractal gas should also be
clustered together. As a result, most stars should be born in clusters
of other stars (Elmegreen et al. 2000). These clusters will then have a
mass spectrum close to $n(M)dM\propto M^{-2}dM$ because hierarchical
structure has equal total mass in equal intervals of log mass. Fractal
structure that is clipped to give only fractal islands has the same mass
spectrum for those islands (Elmegreen 2001). Fractal star fields should
also be correlated with respect to age and distance. Larger regions will
have longer durations of star formation with a duration-size relation
similar to the crossing-time-size relation for interstellar turbulence
(Efremov \& Elmegreen 1998; Battinelli \& Efremov 1999; Harris \&
Zaritsky 1999). Models of this duration-size relation are in Scalo \&
Chappell (1999) and Nomura \& Kamaya (2001).  Evidence for short star
formation times in small regions is in Ballesteros-Paredes, Hartmann,
\& Vazquez-Semadeni (1999) and Elmegreen (2000).

The turbulence model of star formation is important for clusters because
it offers the most natural explanation for the cluster mass spectrum
(Elmegreen \& Efremov 1997), which is observed to be close to $M^{-2}dM$
for local clusters (Battinelli et al. 1994) and OB associations, as seen
by the distribution of HII region luminosities (Kennicutt, Edgar, \&
Hodge 1989; Comeron \& Torra 1996; Rozas, Beckman \& Knapen 1996;
Feinstein 1997; McKee \& Williams 1997; Oey \& Clarke 1998). 

The $\sim M^{-2}dM$ power law also appears at the high mass end of the globular cluster
luminosity function (Ashman, Conti, \& Zepf 1995), and for super-star
clusters in starburst regions (Whitmore \& Schweizer 1995; Zhang \& Fall
1999). 

The turbulence model explains the correlation between the number of
clusters and the mass of the largest cluster (Whitmore 2000) as a size
of sample effect in a random cluster formation model. Similarly, it
explains the correlation between the star formation rate and the
relative fraction of star formation that occurs in the form of clusters
(Larsen \& Richtler 2000). These are both statistical properties for an
ensemble where the clusters form in virialized clouds in a medium with a
uniform average total pressure (total pressure considers the sum of the
ram pressure from flows and the thermal pressure in the compressed
regions). The derivation for the Larsen \& Richtler correlation goes
like this: 

For a general interstellar disk, the total pressure scales with the
product of the gas mass column density and the total mass column density
of all the material, including stars, dark matter, and gas, that lies
inside the gas layer (Elmegreen 1989b). For an interstellar medium that
is close to the threshold for instability, the gas mass column density
is roughly proportional to the total. Thus the pressure scales
approximately as the square of the gas mass column density in a critical
ISM: \begin{equation}P_{ISM}\propto \Sigma_{gas}^2.\end{equation} The
star formation rate scales with this gas column density approximately as
${\rm SFR}\propto \Sigma_{gas}^{1.4}$ (Kennicutt 1998), presumably as a
result of the conversion of the available gas (one power of $\Sigma$)
into stars on a dynamical time scale (the extra fractional power of
$\Sigma$). These two equations give \begin{equation}P_{ISM}\propto {\rm
SFR}^{1.4}.\end{equation}

For clusters, the virial theorem, $c^2\sim0.2GM/R$, and the internal
pressure, $P_{int}\sim0.1GM^2/R^4$ give a relation between mass and
pressure, \begin{equation}M\sim6\times10^3 \;{\rm M}_\odot
\;\left(P_{int}/10^8\;{\rm
K\;cm^{-3}}\right)^{3/2} \left(n/10^5\;{\rm
cm}^{-3}\right)^{-2}.\end{equation} The normalization for this relation
comes from the properties of the molecular core near the Trapezium
cluster in Orion (Lada, Evans \& Falgarone 1997). From this we get
$M\propto P_{int}^{3/2}$ for internal pressure $P_{int}$, which is generally
larger than the environmental pressure, $P_{ISM}$, but proportional to
it on average; thus $M\propto P_{ISM}^{3/2}$. This last step, setting
$P_{int}\propto P_{ISM}$, is more of an assumption than an observation,
but it seems reasonable for an ensemble average. The above equations can
now be combined to give $M\propto {\rm SFR}^2$. 

Now it is important to realize from the fractal model that this mass is
the largest mass than can form, on average, in a virialized cloud at the
ambient pressure. Smaller mass clusters form as part of the
hierarchy of structures on smaller scales. Larger clusters cannot form
systematically because they will be over-pressured. Of course, larger
mass clusters can form in statistical fluctuations of the local
pressure, but the present discussion concerns only the average
properties of clusters. For this reason, we write explicitly
$M_{max}\propto {\rm SFR}^2$ to remind us that this is a maximum cluster
mass. 

The total star formation mass in the form of young dense clusters equals
the integral of the cluster mass weighted by the cluster mass spectrum
over the mass interval ranging from some smallest mass, $M_{min}$, up to
the maximum cluster mass, $M_{max}$. For an $n(M)dM=n_0 M^{-2}dM$
mass spectrum, we first integrate from $M_{max}$ to infinity to 
give the normalization
factor $n_0$, i.e. $\int_{M_{max}}^{\infty}n_0M^{-2}dM=1$, or $n_0=M_{max}$.
Then $n(M)dM=M_{max}M^{-2}dM$. With this normalization, the total mass is
$M_{tot}\propto
M_{max}\ln\left(M_{max}/M_{min}\right)$, which depends only weakly on
$M_{min}$. Considering only the first term, which is the most strongly
varying, we get $M_{tot}\propto M_{max}\propto {\rm SFR}^2$. From this, the fraction of
the star formation in the form of dense clusters, $M_{tot}/{\rm SFR}$,
scales directly with the star formation rate: \begin{equation}
M_{tot}/{\rm SFR}\propto {\rm SFR}\end{equation} This is essentially the
relation found by Larsen \& Richtler (2000). In addition, an
intermediate step in this derivation, not shown here, gives
$M_{max}\propto $ the number of clusters, which is the correlation found
by Whitmore (2000). 

We see from this discussion how easily the turbulence model explains the mass
function of clusters and the correlations between maximum cluster mass,
star formation rate, pressure, and total number of clusters. This means
that turbulence alone explains why starburst regions have more, bigger,
and denser clusters. It may also explain how the high pressure environment
in the early Universe (e.g., in the turbulent proto-halo) led to the
formation of globular clusters. It does not explain the Gaussian
distribution for globular cluster magnitudes, however, which may be a
problem for the model if low mass cluster destruction is so sensitive to
local environment that the theoretical peak of the evolved Gaussian
cluster luminosity function is more variable with environment than the observed peak
(e.g., Vesperini, this conference, but also see Fall, this conference).

\section{Gravitational Instability Model} 

Gravitational instabilities may drive random structure in interstellar
gas, as shown by spiral chaos models (Toomre \& Kaljnas 1991; Huber \&
Pfenniger 1999; Semelin \& Combes 2000) and in the simulations by Wada \& Norman
(1999, 2001) and Wada, Spaans, \& Kim (2000). In these models, gravity
makes scale-free structures because of non-linear interactions between
the primary structures that form at the Jeans length. The Jeans length
is about equal to the scale height, but neither is well defined in a
turbulent medium. 

Gravitational instabilities are easier to see when they occur in spiral
arms and resonance rings. The identification of giant spiral arm cloud
and star-forming complexes as the result of gravitational instabilities has
been made in a series of papers, beginning with Elmegreen (1979) and
including Viallefond, Goss \& Allen (1982), Nakano et al. (1987),
Grabelsky et al. (1987), Elmegreen \& Elmegreen (1983, 1987), Lada et
al. (1988), Ohta, et al. (1988), Boulanger \& Viallefond (1992),
Tilanus \&  Allen (1993), Rand (1993a,b), Garcia-Burillo, Guelin \&
Chernicharo (1993), and Kuno et al. (1995). The identification of giant
star-forming regions in ILR rings as the result of gravitational
instabilities was made by Elmegreen (1994), D. Elmegreen, et al. (1999)
and Buta, Crocker, \& Byrd (1999). Similar complexes in an outer
resonance ring were found in NGC 1300 (Elmegreen et al. 1996). In the
case of ILR rings, there are also super star clusters (Maoz et
al. 1996), perhaps for the reasons given above: the total SFR
is large so the clusters sample far out in their mass function. 

Gravitational instabilities have also shown up in interacting
systems. The merger galaxy NGC 6090 has most of its interstellar medium
between the two nuclei and at very high column density, around
$10^{3.5}$ M$_\odot$ pc$^{-2}$
(Mazzarella \& Boroson 1993; Bryant \& Scoville 1999;
Dinshaw et al. 1999). This gives the average interstellar
pressure a value of $\sim10^{8}$ K cm$^{-3}$, which is four orders of
magnitude larger than the local interstellar pressure. At this position, there
is a chain of four, regularly-spaced, young massive star clusters along
a spiral arm. Their morphology is like the familiar beads-on-a-string
pattern in disk spiral arms (Kuno et al. 1995), but in this case in an
extreme environment. 

Other examples might be the unusual supermassive clusters in NGC 253
(Watson et al. 1996; Keto et al. 1999) and NGC 5253 (Turner, Beck \& Ho
(2000). They are by far the largest clusters in these galaxies. They are
located near the galaxy centers and are not part of any obvious fractal
pattern or sequential triggering event. They are
also not part of a similar chain of clusters, so they do not look like
a beading instability as do complexes in spiral arms or ILR rings.
Nevertheless, they probably formed by gravitational
instabilities in the inner disks of these galaxies. 

\section{Summary} 

Most stars form in clusters and most clusters in the solar neighborhood
formed as a result of direct and sequential triggering stimulated by
other clusters. When viewed from a distance, clusters can have a
fractal and time-correlated pattern suggestive of turbulence. Presumably
ISM turbulence sets up the fractal pattern independent of star
formation, and then star formation inside this pattern operates locally
by a variety of methods, preserving the overall pattern. 
Gravitational instabilities also operate
in the turbulent medium, inside sequentially triggered clouds, and in
larger-scale environments organized by systematic flows. The most
obvious evidence for such organized patterns comes from spiral arms and
resonance rings, where a confinement in two dimensions leads to the
formation of regularly spaced ring hotspots and spiral arm beads on a
string.


\begin{references} 

\reference Asseo, E., Cesarsky, C.J., Lachieze-Ray,
M. \& Pellat, R. 1978, ApJ, 225, L21 

\reference Ashman, K.M., Conti, A., \& Zepf, S.E. 1995, AJ, 110, 1164 

\reference Balbus, S.A., \& Hawley, J.F. 1991, ApJ, 376, 214 

\reference Bally, J., Langer, W.D., Stark, A.A., \& Wilson, R.W. 1987,
ApJ, 312, L45 

\reference Ballesteros-Paredes, J., Hartmann, L., \& Vazquez-Semadeni, E. 1999, 
ApJ, 527, 285

\reference Ballesteros-Paredes, J., Vazquez-Semadeni, E., \& Scalo 1999, 
ApJ, 515, 286

\reference Battinelli P., Brandimarti A., \& Capuzzo-Dolcetta R. 1994,
A\&AS, 104, 379 

\reference Battinelli, P., \& Efremov, Y. N. 1999, A\&A, 346, 778 

\reference Boulanger, F., \& Viallefond, F. 1992, A\&A, 266, 37 

\reference Brinks, E., \& Bajaja, E. 1986, A\&A, 169, 14 

\reference Bryant, P.M., \& Scoville, N.Z.  1999, AJ, 117, 2632

\reference Buta, R., Crocker, D.A., \& Byrd, G.G. 1999, AJ, 118, 2071 

\reference Carpenter, J.M. 2000, AJ, 120, 3139 

\reference Comer\'on, F., \& Torra, J. 1996, A\&A, 314, 776 

\reference Crovisier, J., \& Dickey, J.M. 1983, A\&A, 122, 282 

\reference Dinshaw, N., Evans, A.S., Epps, H., Scoville, N.Z.,
Rieke, M. 1999, ApJ, 525, 702

\reference Dutrey, A., Langer, W.D., Bally, J., Duvert, G., Castets, A.
\& Wilson, R.W. 1991, A\&A, 247, L9 

\reference Efremov, Y. N., \& Elmegreen, B. G. 1998, MNRAS, 299, 588 

\reference Elmegreen, B.G. 1979, ApJ, 231, 372 

\reference Elmegreen, B.G. 1989a, ApJ, 340, 786 

\reference Elmegreen, B.G. 1989b, ApJ, 338, 178 

\reference Elmegreen, B.G. 1993, ApJ, 419, L29 

\reference Elmegreen, B.G. 1994, ApJ, 425, L73 

\reference Elmegreen, B.G. 1998, in Origins, ASP Conf. Series 148, ed.
C.E. Woodward, M. Shull, \& H.A. Thronson, p. 150 

\reference Elmegreen. B.G. 2000, ApJ, 530, 277 

\reference Elmegreen, B.G. 2001, ApJ, in preparation  

\reference Elmegreen, B.G., \& Elmegreen, D.M. 1983, MNRAS, 203, 31 

\reference Elmegreen, B.G., \& Elmegreen, D.M. 1987, ApJ, 320, 182 

\reference Elmegreen, B.G., \& Efremov, Yu. N. 1997, ApJ, 480, 235 

\reference Elmegreen, B.G., Efremov, Y.N., Pudritz, R., \& Zinnecker, H.
2000, in Protostars and Planets IV, ed. V. G. Mannings, A. P. Boss, \&
S. S. Russell (Tucson: Univ. Arizona Press), 179 

\reference Elmegreen, B.G., Kim, S., \& Staveley-Smith, L. 2001, ApJ,
548, 749 

\reference Elmegreen, B.G., \& Elmegreen, D.M. 2001, AJ, 121, 1507 

\reference Elmegreen, D.M., Chromey, F., Elmegreen, B.G., \&
Hasselbacher, D. 1996, ApJ, 469, 131 

\reference Elmegreen, D.M., Chromey, F.R., Sawyer, J.E., \& Reinfeld,
E.L. 1999, AJ, 118, 777 

\reference Evans, N.J., II, 1999, ARAA, 37, 311 

\reference Feinstein, C. 1997, ApJS, 112, 29 

\reference Garcia-Burillo, S., Guelin, M., \& Chernicharo, J. 1993,
A\&A, 274, 123 

\reference de Geus, E.J. 1992, A\&A, 262, 258 

\reference Gomez, M., Hartmann, L., Kenyon, S. J., \& Hewett, R. 1993,
AJ, 105, 1927 

\reference Grabelsky, D.A., Cohen, R.S., May, J., Bronfman, L., \&
Thaddeus, P. 1987, ApJ, 315, 122 

\reference Green, D.A. 1993, MNRAS, 262, 327 

\reference Harris, J., \& Zaritsky, D. 1999, AJ, 117, 2831 

\reference Huber, D., \& Pfenniger, D. 1999, in The Evolution of
Galaxies on Cosmological Timescale, eds. J.E. Beckman \& T.J. Mahoney,
Astrophysics and Space Science, poster paper

\reference Kennicutt, R.C., Jr. 1998, ApJ, 498, 541 

\reference Kennicutt, R.C., Edgar, B.K., \& Hodge, P.W. 1989, ApJ, 337,
761 

\reference Kerton, C.R., \& Martin, P.G. 2000, ApJS, 126, 85 

\reference Keto, E., Hora, J.L., Fazio, G.G., Hoffmann, W., \& Deutsch,
L. 1999, ApJ, 518, 183 

\reference Klessen, R.S., Heitsch, F., \& Mac Low, M.-M. 2000, 535, 887 

\reference Kuno, N., Nakai, N., Handa, T., \&  Sofue, Y. 1995, PASJ, 47, 745

\reference Lada, C.J., Margulis, M., Sofue, Y., Nakai, N., \& Handa, T.
1988, ApJ, 328, 143 

\reference Lada, C. J., Depoy, D. L., Merrill, K. M., \& Gatley, I.
1991, ApJ, 374, 533 

\reference Lada, C.J., Alves, J., \& Lada, E.A. 1996, AJ, 111, 1964 

\reference Lada, C.J., Alves, J., \& Lada, E.A. 1999, ApJ, 512, 250 

\reference Lada, E.A., \& Lada, C.J. 1995, AJ, 109, 1682 

\reference Lada, E.A., Evans, N.J., II., \& Falgarone, E. 1997, ApJ,
488, 286 

\reference Larsen, S.S., \& Richtler, T. 2000, A\&A, 354, 836 

\reference Lazarian, A., \& Pogosyan, D. 2000, ApJ, 537, 720 

\reference Lefloch, B., \& Cernicharo, J. 2000, ApJ, 545, 340 

\reference MacLow, M.-M., Klessen, R. S., Burkert, A., \& Smith, M. D.
1998, Phys. Rev. Lett., 80, 2754 

\reference Maoz, D., Barth, A.J., Sternberg, A., Filippenko, A.V., Ho,
L.C., Macchetto, F.D., Rix, H.W., \& Schneider, D.P. 1996, AJ, 111, 2248

\reference Mazzarella, J.M., \& Boroson, T.A. 1993, ApJS, 85, 27

\reference McCray, R., \& Kafatos, M. 1987, ApJ, 317, 190 

\reference McKee, C.F., \& Williams, J.P. 1997, 476, 144 

\reference Nakano, M., Ichikawa, T., Tanaka, Y.K., Nakai, N., \& Sofue,
Y. 1987, PASJ, 39, 57 

\reference Nomura, H., \& Kamaya, H. 2001, AJ, 121, 1024 

\reference Oey, M. S., \& Clarke, C. J. 1998, AJ, 115, 1543 

\reference Ohta, K., Sasaki, M., \& Saito, M. 1988, PASJ, 40, 653 

\reference Padoan, P., Juvela, M., Goodman, A.A., \& Nordlund, A. 2001,
ApJ, 553, 227

\reference Palla, F., \& Stahler, S.W. 1999, ApJ, 525, 772 

\reference Palla, F., \& Stahler, S.W. 2000, ApJ, 540, 255 

\reference Parker, E.N. 1966, ApJ, 145, 811 

\reference Phelps, R.L., \& Lada, E.A. 1997, ApJ, 477, 176 

\reference Preibisch, T., \& Zinnecker, H. 1999, AJ, 117, 2381 

\reference Puche, D., Westpfahl, D., Brinks, E., \& Roy, J-R. 1992, AJ,
103, 1841 

\reference Rand, R.J. 1993a, ApJ, 404, 593 

\reference Rand, R.J. 1993b, ApJ, 410, 68 

\reference Reipurth, B., Rodriguez, L.F., \& Chini, R. 1999, AJ, 118, 983 

\reference Reynoso, E.M., \& Mangum, J.G. 2001, AJ, 121, 347

\reference Rozas, M., Beckman, J.E., \& Knapen, J. H. 1996, 307, 735

\reference Sancisi, R., Goss, W. M., Anderson, C., Johansson, L. E. B., \& Winnberg, A. 1974,
A\&A, 35, 445

\reference Sargent, A.I. 1979, ApJ, 233, 163 

\reference Scalo, J., \& Chappell, D. 1999, ApJ, 510, 258

\reference Sellwood, J.A., \& Balbus, S.A. 1999, ApJ, 511, 660 

\reference Semelin, B., \& Combes, F. 2000, A\&A, 360, 1096

\reference Stanimirovic, S., Staveley-Smith, L., Dickey, J.M., Sault,
R.J., \& Snowden, S.L. 1999, MNRAS, 302, 417 

\reference Stewart, S.G., Fanelli, M.N., Byrd, G.G., Hill, J.K., Westpfahl, D.J.,
Cheng, K.-P., O'Connell, R.W., Roberts, M.S., Neff, S.G., Smith, A.M., \&
Stecher, T.P. 2000, ApJ, 529. 201

\reference Stewart, S.G., \& Walter, F. 2000, AJ, 120, 1794 

\reference Stone, J.M., Ostriker, E.C., \& Gammie, C.F. 1998, ApJ, 508,
99 

\reference St\"utzki, J., Bensch, F., Heithausen, A., Ossenkopf, V., \&
Zielinsky, M. 1998, A\&A, 336, 697 

\reference Testi, L., Sargent, A.I., Olmi, L., \& Onello, J.S. 2000,
ApJ, 540, L53 

\reference Tilanus, R.P.J., \& Allen, R.J., 1993, A\&A, 274, 70 

\reference Toomre, A., \& Kalnajs, A.J. 1991, in Dynamics of Disk
Galaxies, ed. B. Sundelius, University of Chalmers, p. 341 

\reference Turner, J.L., Beck, S.C., \& Ho, P.T.P. 2000, ApJ, 532, L109 

\reference V\'azquez-Semadeni, E., Passot, T., \& Pouquet, A. 1996, ApJ,
473, 881 

\reference Viallefond, F., Goss, W.M., \& Allen, R.J. 1982, A\&A, 115,
373 

\reference Wada, K., \& Norman, C. A. 1999, ApJ, 516, L13 

\reference Wada, K., \& Norman, C. A. 2001, ApJ, 547, 172

\reference Wada, K., Spaans, M., \& Kim, S. 2000, ApJ, 540, 797 

\reference Walter, F., \& Brinks, E. 1999, AJ, 118, 273 

\reference Watson et al. 1996 AJ 112, 534 

\reference Westpfahl, D.J., Coleman, P.H., Alexander, J., \& Tongue, T.
1999, AJ, 117, 868 

\reference Whitmore, B.C. 2000, in STScI Symposium Series 14, ed. M.
Livio, astroph/0012546 

\reference Whitmore, B.C., \& Schweizer, F. 1995, AJ, 109, 960 

\reference Wilcots, E.M, \& Miller , B.W. 1998, AJ, 116, 2363

\reference Xie, T., \& Goldsmith, P.F. 1994, ApJ, 430, 252 

\reference Yamaguchi, R., Mizuno, N., Onishi, T., Mizuno, A., \& Fukui,
Y. 2001, ApJL, 553, 185

\reference Zhang, Q., \& Fall, S.M. 1999, ApJ, 527, L81 

\end{references}
\end{document}